\documentclass[a4paper,UKenglish,cleveref, autoref, thm-restate]{lipics-v2021}

\pdfoutput=1 
\hideLIPIcs  

\title{A Generalized Quantum Branching Program} 

\author{Debajyoti Bera}{Department of Computer Science, IIIT-D, New Delhi, India }{dbera@iiitd.ac.in}{}{}

\author{Tharrmashastha SAPV}{Department of Computer Science, IIIT-D, New Delhi, India}{tharrmashasthav@iiitd.ac.in}{}{}

\authorrunning{D. Bera and Tharrmashastha SAPV} 

\Copyright{Debajyoti Bera and Tharrmashastha SAPV} 



\category{} 

\relatedversion{} 

\nolinenumbers 

\EventEditors{John Q. Open and Joan R. Access}
\EventNoEds{2}
\EventLongTitle{42nd Conference on Very Important Topics (CVIT 2016)}
\EventShortTitle{CVIT 2016}
\EventAcronym{CVIT}
\EventYear{2016}
\EventDate{December 24--27, 2016}
\EventLocation{Little Whinging, United Kingdom}
\EventLogo{}
\SeriesVolume{42}
\ArticleNo{23}

\usepackage{physics}
\usepackage{xcolor}
\usepackage{xspace}
\usepackage{todonotes}
\usepackage{tikz}
\usepackage{graphicx}
\usepackage{float}

\newcommand{\nqbp}{NQBP\xspace}
\newcommand{\aqbp}{AQBP\xspace}
\newcommand{\obdd}{OBDD\xspace}
\newcommand{\qobdd}{QOBDD\xspace}
\newcommand{\nqobdd}{NQOBDD\xspace}

\newcommand{\gqbp}{GQBP\xspace}
\newcommand{\pbp}{PBP\xspace}
\newcommand{\what}[1]{\widehat{#1}}

\newcommand{\scra}{$\mathcal{A}$\xspace}  
\newcommand{\scrg}{$\mathcal{G}$\xspace}  
\newcommand{\aqbpcl}[2]{\texttt{AQBP}_{#1,#2}} 
\newcommand{\gqbpcl}[2]{\texttt{GQBP}_{#1,#2}} 

\begin{document}

\maketitle

\begin{abstract}
Classical branching programs are studied to understand the space complexity of computational problems. Prior to this work, Nakanishi and Ablayev had separately defined two different quantum versions of branching programs that we refer to as NQBP and AQBP. However, none of them, to our satisfaction, captures the intuitive idea of being able to query different variables in superposition in one step of a branching program traversal. Here we propose a quantum branching program model, referred to as GQBP, with that ability. To motivate our definition, we explicitly give examples of GQBP for n-bit Deutsch-Jozsa, n-bit Parity, and 3-bit Majority with optimal lengths. We the show several equivalences, namely, between GQBP and AQBP, GQBP and NQBP, and GQBP and query complexities (using either oracle gates and a QRAM to query input bits). In way this unifies the different results that we have for the two earlier branching programs, and also connects them to query complexity. We hope that GQBP can be used to prove space and space-time lower bounds for quantum solutions to combinatorial problems.
\end{abstract}

\section{Introduction}
The currently popular NISQ model of quantum computing highlights the importance of quantum space, i.e., qubit usage of quantum algorithms. Motivated by the need to study the quantum space complexity of combinatorial problems, we naturally looked towards quantum branching programs. A branching program is like a ``random access finite-state machine where only one input character can be looked up in any state''~\cite{masek}. In the classical domain, branching programs captures the non-uniform space complexity of a problem in a very natural way~\cite{razborov1991lower}. A branching program of size $s$, i.e., with $s$ states, can be simulated by any (non-uniform) model with $\log(s)$ space to store the state space. Similarly, if a computation involves $c$ distinct internal states or configurations (e.g., a $\log(c)$-space-bounded Turing machine), it is possible to simulate the computation by a $c$-size branching program. So, e.g., polynomial-size branching programs are equivalent to the non-uniform class $LOGSPACE/poly$.

Our interest in quantum branching programs is also related to a ``more practical'' motivation related to a current trend in machine learning --- that of explainability. A decision tree is arguably the most intuitive machine learning model that can be explained in a straightforward manner. Masek, in his Master's thesis, studied branching programs in the name of decision graphs, which are essentially compact forms of decision trees. We believe that quantum branching programs may be a worthy machine learning model that can also be explained.

We are aware of two prior attempts to define a quantum version of branching programs. They, as well as ours, follow the general idea of changing the current state (which could be a superposition of basis states) by querying some input bit; the difference primarily lies in how evolution happens. In the first version defined by Nakanishi et al.~\cite{nakanishi2000ordered} (we refer to this as \nqbp), the entire state is measured after each step, and evolution is continued if neither an accept nor a reject state is observed. In the version defined by Ablayev et al.~\cite{ablayev2001computational}, in each step of the evolution, a fixed bit is queried; we refer to these programs as \aqbp).

Quite a few results are known for these branching programs. The first advantage of quantum branching programs, more specifically \nqbp, was demonstrated by Nakanishi et al. in~\cite{nakanishi2000ordered}. 
They defined a function, namely \textit{Half Hamming weight function}, that checks if the Hamming weight of the input is exactly $n/2$.
They showed that there exists an ordered bounded width \nqbp (also known as an \nqobdd) that can compute the half Hamming weight function while no ordered
bounded width PBP can compute that function.

This was followed by the work of Ablayev et al~\cite{ablayev2001computational} in which they demonstrated that an $(O(\log p_n), n)$ \aqbp~can compute the
$Mod_{p_n}$ function with one-sided error. However, any stable probabilistic \obdd~needs width at least $p_n$.
Here, $Mod_{p_n}$ function is the function that decides if the Hamming weight of the input is divisible by a prime $p_n$.
They also gave the seminal result that the class of languages in $NC1$, the languages that are decidable by uniform Boolean circuits of $O(\log n)$ depth and polynomial number of gates with at most two inputs, lies within the class of languages decidable by width-2 AQBPs~\cite{ablayev2002quantum}.

Later, Sauerhoff et al. in~\cite{sauerhoff2005quantum} proved two results that exhibit the power and limitation of \nqbp. 
First, they showed that the \textit{permutation matrix test function}, the function that tests if the given matrix is a permutation matrix, can be computed using a small sized N\qobdd~but needs a large deterministic~\obdd.
Alternatively, they also proved that any N\qobdd~that computes the \textit{disjointness function} requires $O(2^{\Omega(n)})$ size, but there exists a small sized deterministic \obdd~that computes the function.
More recently, in~\cite{maslov2021quantum}, Maslov et al. demonstrated that any symmetric Boolean function can be computed by $(2, O(n^2))$-\aqbp.
This is a striking result because there are symmetric functions that need at least $3$ bits of computational space to be computed in polynomial time with arbitrary error and symmetric functions with non-trivial Fourier spectrum are not computable in polynomial time.
Observe that a trivial lower bound on \aqbp for computing any non-constant symmetric function is $\Omega(n)$ because the function is oblivious to the exact input and depends only on the Hamming weight of the input.

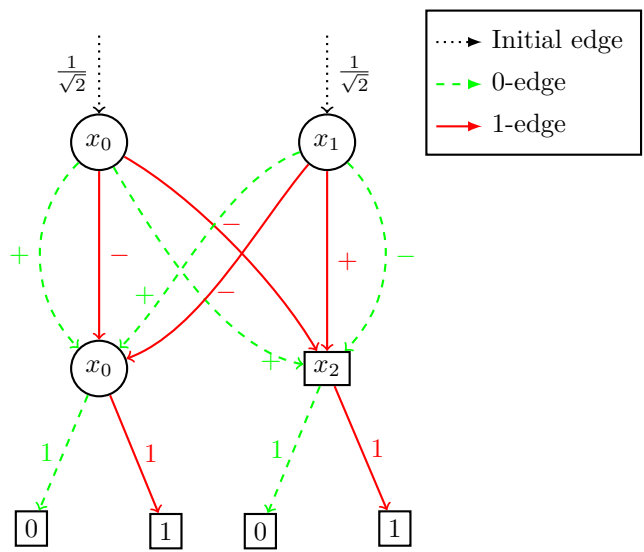
\begin{figure}[ht!]
\centering
\label{fig:maj-3}
        \begin{tikzpicture}[scale=0.5, node distance={30mm}, thick, main/.style = {draw, circle}, empty/.style = {circle}, rect/.style={draw, rectangle}] 
            \node[main] (0) {$x_0$}; 
            \node[main] (1) [right of=0] {$x_1$};
            \node[main] (2) [below of=0] {$x_0$};
            \node[rect] (3) [right of=2]{$x_2$}; 
            \node[rect] (4) [below left of=2, xshift=35pt] {$0$};
            \node[rect] (5) [below of=2, yshift=24pt, xshift=25pt] {$1$};
            \node[rect] (6) [below of=3, yshift=24pt, xshift=-25pt] {$0$};
            \node[rect] (7) [below right of=3, xshift=-35pt] {$1$};
            \node[empty](8) [above of=0, yshift=-40pt] {};
            \node[empty](9) [above of=1, yshift=-40pt] {};

            \draw[->, dotted] (8) to node[left] {$\frac{1}{\sqrt{2}}$} (0);
            \draw[->, dotted] (9) to node[right] {$\frac{1}{\sqrt{2}}$} (1);
            
            \draw[->, green, dashed] (0) to [out=-135,in=135,looseness=1] node[left] {+} (2);
            \draw[->, red] (0) to node[right] {$-$} (2);
            \draw[->, green, dashed] (0) to [out=-60,in=170,looseness=0.75] node[below right=30pt] {+} (3);
            \draw[->, red] (0) to [out=-30,in=120,looseness=0.75] node[above] {$-$} (3);

            \draw[->, green, dashed] (1) to [out=200,in=45,looseness=0.75] node[below left=20pt] {+} (2);
            \draw[->, red] (1) to [out=230,in=20,looseness=0.75] node[below] {$-$} (2);
            \draw[->, green, dashed] (1) to [out=-45,in=45,looseness=1] node[right] {$-$} (3);
            \draw[->, red] (1) to node[right] {+} (3);

            \draw[->, green, dashed] (2) to node[left] {1} (4);
            \draw[->, red] (2) to node[right] {1} (5);

            \draw[->, green, dashed] (3) to node[left] {1} (6);
            \draw[->, red] (3) to node[right] {1} (7);

            \matrix [draw,below right] at (current bounding box.north east) {
                \draw[-latex,color=black, dotted](0,-0.3) -- (0.6,-0.3); & \node{Initial edge};\\
                \draw[-latex,color=green, dashed](0,-0.3) -- (0.6,-0.3); & \node{$0$-edge};\\
                \draw[-latex,color=red](0,-0.3) -- (0.6,-0.3); & \node{$1$-edge};\\
            };
            
        \end{tikzpicture} 
        \caption{\label{fig:gqbp-maj3} A length-2 and width-4 \gqbp to (exactly) compute the majority of 3 bits $x_0,x_1,x_2$. Here, + and $-$ denotes the amplitudes $\tfrac{1}{\sqrt{2}}$ and $-\tfrac{1}{\sqrt{2}}$, respectively.
        }
    \end{figure}

Despite the above encouraging results, we observed that \aqbp suffers from the restriction of querying the same input bit within the different superposition paths in any single level. Further, the idea of measuring after every evolution step in an \nqbp was not a very agreeable one. Therefore, we defined a new version which we refer to as \gqbp (Figure~\ref{fig:gqbp-maj3} shows an example) that combines the idea of completely independent superposition paths as in \nqbp with the idea of performing one measurement right at the end (as in \aqbp), thus allowing interference to show its might. We use the complexity measures width (to denote the largest number of states or nodes in any level) and length (to denote the length of the longest path in the state diagram. Even though we do not consider size as a measure in this work, our results can be interpreted in terms of size by using the trivial upper bound of $size \le width \times length$.


This paper introduces the \gqbp model of branching program and relates it to a few other models like AQBP, NQBP, and query complexity. We are able to show the following.

\begin{itemize}
    \item We show explicit general branching programs for {\em Parity} of $n$-bits, the problem of {\em Deutsch-Jozsa}, and {\em Majority} of 3 bits. These branching programs are optimal with respect to their lengths, and the optimality follows from a relation between the length of the shortest \gqbp to the polynomial degree of a problem.

    \item We prove that an \aqbp is a special form of \gqbp (in which the same variable is queried in each level), and thus, the former cannot compute functions that depend non-trivially on every input bit using less than $n$ length. We present a \gqbp for Parity of length $n/2$, and this shows that \gqbp does not suffer from the same limitation and is strictly more powerful, especially in the sub-linear length regime. 
    
    \item Due to the above point, all the results concerning \aqbp now also hold for \gqbp, e.g., any problem in $NC^1$ can now be solved using a width-2 \gqbp, and any symmetric Boolean function can be computed using a width-2, $O(n^2)$-length \gqbp.
    

    \item Despite their structural difference, it is not difficult to show that a \gqbp can be viewed as an \nqbp. We further show that an \nqbp can be simulated by a \gqbp without any overhead of size and length.

    \item Next, we turn to quantum circuits. First, we show how to design an \aqbp (and hence a \gqbp) with width $2^q$ and length $t$ to simulate a quantum query circuit in the QRAM model making $q$ queries and consuming $t$ qubits. The QRAM query model provides an alternative to implementing the oracle (in the oracle query model) by having the input in query-only registers ({\it aka.} QRAM) that can be queried using CNOT (or similar) gates. We want to point out that even though it may be possible to obtain a \gqbp for $Parity_n$ using this approach, our handcrafted \gqbp for the same uses lesser width.

    \item Next, we show the reverse direction, i.e., how to design a quantum circuit (using either an oracle or a QRAM to access input bits) for a given \gqbp using reasonably tight depth and space (i.e., qubits). This and the above point allows us to interpret the upper and the lower bounds in quantum query complexities in terms of a \gqbp and possibly derive new insights into quantum complexities of problems.
\end{itemize}

\section{Background: Quantum Branching Programs}
\label{sec:qbp-background}

In this section, we familiarize the reader with the existing branching program frameworks. We then introduce a new variant of quantum branching programs.

The most basic branching programs are the deterministic branching programs which are defined as follows.

\begin{definition}[Deterministic Branching Program]
    A deterministic branching program is a tuple $(Q, E, v_0, L, \delta, F)$ along with an evolution operator $E$ where
    \begin{itemize}
        \item $Q$ is a set of nodes.
        \item $E$ is the set of edges on $Q$ such that $(Q,E)$ form a directed acyclic multi-graph.
        \item $L:Q \xrightarrow{} \{1,2,\cdots, n\}$ assigns a variable to each node.
        \item $v_0\in Q$ is the start node.
        \item $\delta: Q\otimes \{0,1\}\xrightarrow{} Q$ is transition function.
        \item $F\subseteq Q$ is the set of sinks such that $F=F_a \cup F_r$ where $F_a$ and $F_b$ are the accept and reject sinks respectively.
    \end{itemize}
\end{definition}
Given an input $x$, starting from the node $v_0$, one iteration of the evolution of the deterministic branching program consists of two steps:
\begin{enumerate}
    \item Let the current node be $v$. Query the value of $x_{L(v)}$ and call it $j$. Move to the node $\delta(v, j) = v'$.
    \item If $v' \in F_a$ then accept the input, else if $v'\in F_b$ then reject the input. Else, continue.
\end{enumerate}

The complexity of a branching program is primarily categorized by two measures: \textit{size} and \textit{length} of a branching program.
\begin{definition}[Size and length of a branching program]
    \begin{itemize}
        \item The \textbf{size} of a branching program is the total number of nodes in the branching program.
        \item The \textbf{length} of a branching program is the length of the longest path from the start node to any sink node.
    \end{itemize}
\end{definition}

It is well known that any classical branching program can be transformed into an equivalent levelled branching program, i.e., a branching program where all the transitions are such that they go from a node at a level $i$ only to a node at level $i+1$ for each $i$.
From now on, we assume that the branching program is levelled unless otherwise stated.

For levelled branching program, another well-used measure of complexity is the \textit{width} of a branching program.
\begin{definition}[width of a leveled branching program]
    The \textbf{width} of a levelled branching program is the maximum number of nodes in any level of the branching program.
\end{definition}

A straightforward generalization of the deterministic branching programs is the probabilistic branching programs. 
But before we define the probabilistic branching programs, we establish well-behaved transition functions.
A transition function $\delta: Q\otimes \{0,1\}\times Q\xrightarrow{} \mathbb{R}$ is said to be \textbf{classically well-behaved} if for any $v\in Q$ and $a\in \{0,1\}$,
\begin{equation*}
    \sum_{v'\in Q}\delta(v, a, v') = 1.
\end{equation*}

Now, the probabilistic branching programs are defined as below.

\begin{definition}[Probabilistic Branching Program]
    A probabilistic branching program is a tuple $(Q, E, v_0, L, \delta, F)$ along with an evolution operator $E$ where
    \begin{itemize}
        \item $Q$ is a set of nodes.
        \item $E$ is the set of edges on $Q$ such that $(Q,E)$ form a directed acyclic multi-graph.
        \item $L:Q \xrightarrow{} \{1,2,\cdots, n\}$ assigns a variable to each node.
        \item $v_0\in Q$ is the start node.
        \item $\delta: Q\otimes \{0,1\}\times Q\xrightarrow{} \mathbb{R}$ is a classically well-behaved transition function.
        \item $F\subseteq Q$ is the set of sinks such that $F=F_a \cup F_r$ where $F_a$ and $F_b$ are the accept and reject sinks respectively.
    \end{itemize}
\end{definition}

Given an input $x$, starting from the node $v_0$, one iteration of the evolution of the deterministic branching program consists of two steps:
\begin{enumerate}
    \item Let the current node be $v$. Query the value of $x_{L(v)}$ and call it $j$. Move to the node $v'$ with probability $\delta(v, j, v')$.
    \item If $v' \in F_a$ then accept the input, else if $v'\in F_b$ then reject the input. Else, continue.
\end{enumerate}

An input string is {\em accepted} by a branching program if the evolution ends in an accept state. 
In this paper, we are concerned with {\em exact computation}, i.e., the probability of acceptance should be 1 for good inputs and 0 for bad inputs.

In general, branching programs are complex structures to analyse. For ease of analysis, various variants of the branching programs were introduced. Some of the variants are as follows.
\begin{itemize}
    \item \textbf{Oblivious branching programs : } These are levelled branching programs in which the variable to be queried depends on the level of the current node rather than the node itself.
    \item \textbf{Ordered binary decision diagrams (OBDD) : } These are branching programs with a fixed ordering of querying the indices along any path.
    \item \textbf{Read-k branching programs : } These are branching programs in which the number of times any variable is queried is at most $k$ along any path from the source node to a sink node.
\end{itemize}

The notion of quantum branching programs was first introduced by Nakanishi et al.\ in~\cite{nakanishi2000ordered} as an extension of the classical probabilistic branching programs.

\begin{definition}[Nakanishi's Quantum Branching Program (\nqbp)]
    A quantum branching program is a tuple $(Q,E,v_0,L,\delta, O)$ where
    \begin{itemize}
        \item $Q = Q_{acc} \cup Q_{rej} \cup Q_{non}$ is a set of nodes.
        \item $E$ is the set of edges on $Q$ such that $(Q,E)$ is a directed multi-graph.
        \item  $v_0$ is the start node.
        \item $L:Q \xrightarrow{} \{1,2,\cdots, n\}$ assigns a variable to each node.
        \item $\delta:Q\otimes \{0,1\}\otimes Q \xrightarrow{} \mathbb{C}$ is a quantumly well-behaved transition function.
        \item $O = \ketbra{E_{acc}} + \ketbra{E_{rej}} + \ketbra{E_{non}}$ where 
        \begin{itemize}
            \item $\ket{E_{acc}} = span\{\ket{v} : v\in Q_{acc}\}$
            \item $\ket{E_{rej}} = span\{\ket{v} : v\in Q_{rej}\}$
            \item $\ket{E_{non}} = span\{\ket{v} : v\in Q_{non}\}$
        \end{itemize}
    \end{itemize}
\end{definition}

Given an oracle $O_x$ for the input, the $i^{th}$ iteration of the evolution of \nqbp~comprises of the following steps:
\begin{enumerate}
    \item Define an operator $U^O:Q \xrightarrow{} Q$ as 
    \begin{equation*}
        U^{O_x}\ket{v} = \sum_{v'\in Q}\delta(v,O_x(L(v)),v')\ket{v'}.
    \end{equation*}
    \item Apply $U^{O_x}$ on $\ket{\psi_i} = U^{O_x}\ket{\psi_{i-1}}$, $\ket{\psi_0} = \ket{v_0}$.
    \item Observe $\ket{\psi_i}$ with respect to $O$.
    \item If the post-measurement state is in the subspace spanned by $\ket{E_{non}}$, continue. Else accept or reject depending on the outcome.
\end{enumerate}

A transition function $\delta:Q\otimes \{0,1\}\otimes Q \xrightarrow{} \mathbb{C}$ is said to be \textbf{quantumly well-behaved} if for any $v_1, v_2\in Q$ and all inputs $x\in \{0,1\}^n$,
\begin{equation*}
    \sum_{v'\in Q}\delta(v_1, x_{L(v_1)}, v')\delta^*(v_2, x_{L(v_2)}, v') = \begin{cases}
        1~\text{ if }~v_1=v_2\\
        0~\text{ if }~v_1\neq v_2\\
    \end{cases}
\end{equation*}

Closely followed by this work, Ablayev et al., in \cite{ablayev2001computational}, proposed a new model for quantum branching programs that was significantly different from \nqbp s.
Before we define Ablayev's model, we first define a quantum transformation as defined by Ablayev et al.

\begin{definition}[Quantum Transformation]
    Given an input string $x\in \{0,1\}^n$, a $d$-dimensional quantum transformation $\langle j, U(0), U(1)\rangle$ on a state $\ket{\psi}\in \mathcal{H}^{\otimes d}$ is defined as
    \begin{equation*}
        \ket{\psi'} = U(x_j)\ket{\psi}.
    \end{equation*}
    where $U(0)$ and $U(1)$ are a $d\times d$ unitary matrices.
    Intuitively, the quantum transformation applies $U(0)$ to $\ket{\psi}$ if the $j^{th}$ bit of $x$ is $0$ and applies $U(1)$ if the $j^{th}$ bit of $x$ is $1$.
\end{definition}

Now, we define Ablayev's quantum branching program model.

\begin{definition}[Ablayev's Quantum Branching Programs (\aqbp)]
    A quantum branching program $P$ of width $d$ and length $l$ is a tuple $\langle T, \ket{\psi_0}, F\rangle$ where
    \begin{itemize}
        \item $T = \big( \langle j_i, U_i(0), U_i(1) \rangle \big)_{i=1}^l$ is a sequence of $d$-dimensional quantum transformations.
        \item $\ket{\psi_0}$ is the initial configuration of $P$.
        \item $F\subseteq \{\ket{0},\ket{1},\cdots, \ket{d-1}\}$ is the set of accepting states.
    \end{itemize}
\end{definition}

Given an input $x = x_1x_2\cdots x_n$, the \aqbp~$P$ transforms as 
\begin{equation*}
    \ket{\psi_x} = U_l(x_{j_l}) U_{l-1}(x_{j_{l-1}})\cdots U_2(x_{j_2})U_1(x_{j_1})\ket{\Psi_0}.
\end{equation*}
To obtain the output, the state $\ket{\psi_x}$ is measured with projection matrix $M$ such that $M_{ii}=1$ iff $i\in F$ and $0$ else.
The probability that $P$ accepts the input $x$ is 
\begin{equation*}
    P_{accept}(x) = ||M\ket{\psi_x}||^2
\end{equation*}

Notice that this model is the quantum equivalent of the oblivious classical branching programs; the query made at any given level is independent of the nodes in that level of the \aqbp.

We define a new model of branching programs that differs from the above two models.

\begin{definition}[Generalized quantum branching program (\gqbp)]
    A quantum branching program is a tuple $(Q, E, \ket{v_0}, L, \delta,  F)$ where
    \begin{itemize}
        \item $Q = \bigcup_{i=0}^{l} Q_i$ is a set of nodes where $Q_i$ is the set of nodes at level $i$.
        \item $E$ is a set of edges between the nodes in $Q$ such that $(Q,E)$ is a levelled directed acyclic multi-graph.
        \item $\ket{v_0}$ is the initial state which is a superposition of the nodes in $Q_0$.
        \item $\delta : Q\times \{0,1\} \times Q \xrightarrow{} \mathbb{C}$ is a \textbf{quantumly well-behaved} transition function.
        \item $L : Q \xrightarrow{} \{1,2\cdots, n\}$ is a function that assigns a variable to each node in $Q\setminus F$.
        \item $F\subseteq Q$ is the set of terminal nodes.
    \end{itemize}
\end{definition}

The evolution of this \gqbp is defined as follows:
\begin{enumerate}
    \item Given access to an input $x$ using oracle $O_x$, define $U^{O_x}_i : \mathcal{H}(d)\xrightarrow{}\mathcal{H}(d)$ for $i\in \{1,2,\cdots, l\}$ that has query access to $O_x$ as
    \begin{equation*}
        U^{O_x}_i \ket{v} =  \sum_{v'\in Q_i}\delta(v, O_x(L(v)), v')\ket{v'}
    \end{equation*}
    for any $\ket{v}\in Q_{i-1}$ where $\mathcal{H}(d)$ is the Hilbert space on $d$ dimensions with $d$ as the width of the QBP.
    \item  Start at $\ket{v_0}$.
    \item For $i=1$ to $l$, perform
    \begin{equation*}
        \ket{\psi_{i}} = U^{O_x}_{i}\ket{\psi_{i-1}}
    \end{equation*}
    where $\ket{\psi_0} = \ket{v_0}$.
    \item Measure $\ket{\psi_{l}}$ in the standard basis as $m$. If $\ket{m}\in F$ then accept the input. Else, reject the input.
\end{enumerate}


It is quite straightforward that this model of quantum branching program subsumes the \aqbp~model.
So, the set of languages computable by \aqbp s is a subset of that computable by \gqbp s.

\section{Upper Bounds on \gqbp}
\label{sec:upper-bounds}

In this section, we show some GQBPs for some simple Boolean functions.
We first start with the classic Deutsch-Josza problem, which is a promise problem whose task is to identify if an $n$-bit binary string is balanced (has Hamming weight $n/2$) or is constant (has Hamming weight $0$ or $n$).

\begin{theorem}
    $DJ_n \in \gqbpcl{2n}{1}$.
\end{theorem}
\begin{proof}
    Consider the two-layered \gqbp \scrg$=(Q,E,\ket{v_0},L,\delta,F)$ where $Q = Q_0\cap Q_1$, $Q_i=\{v_{i,j}: j\in \{1,\cdots, n\}\}$, $E = \{(v_{0,i}, v_{1,j}): i,j\in \{1,\cdots, n\}\}$ and $F = Q_1\setminus \{\ket{v_{1,1}}\}$.
    Define the function $L$ as $L(v_{0,i}) = x_i$ for $i\in \{1,\cdots, n\}$.
    Also, define the transition function $\delta$ as 
    \begin{equation*}
        \delta(v_{0,i}, a, v_{1,j}) = \frac{1}{2^{n/2}}(-1)^a (-1)^{i\cdot j}
    \end{equation*}
    for all $i,j \in \{1,\cdots, n\}$.
    Now, set $\ket{v_0} = \frac{1}{2^{n/2}}\sum_{v_{0,i}\in Q_0}\ket{v_{0,i}}$.
    Fix an input $x$. Then the transition unitary $U_{1}^{O_x}$ will be defined as 
    \begin{equation*}
        U_{0}^{O_x}\ket{v_{0,i}} = \sum_{v_{1,j}\in Q_1} \delta(v_{0,i}, x_i, v_{1,j})\ket{v_{1,j}} = \frac{1}{2^{n/2}}\sum_{v_{1,j}\in Q_1}(-1)^{x_i}(-1)^{i\cdot j} \ket{v_{1,j}}.
    \end{equation*}
    Then the final state of \scrg before measurement can be calculated as
    \begin{align*}
        U_{0}^{O_x}\ket{v_{0}} &= \frac{1}{2^n}\sum_{v_{0,i}\in Q_0}U_{0}^{O_x}\ket{v_{0,i}}\\
        &= \frac{1}{2^n}\sum_{v_{0,i}\in Q_0} \frac{1}{2^{n/2}}\sum_{v_{1,j}\in Q_1}(-1)^{x_i}(-1)^{i\cdot j} \ket{v_{1,j}}\\
        &= \frac{1}{2^n} \sum_{v_{1,j}\in Q_1}\Big[\sum_{v_{0,i}\in Q_0} (-1)^{x_i \oplus i\cdot j}\Big] \ket{v_{1,j}}
    \end{align*}
    Since, $F = Q_1\setminus \{\ket{v_{1,1}}\}$, the probability of accepting the input $x$ is 
    \begin{equation*}
        Pr_\mathcal{G}[accept] = 1-\Bigg\lvert \sum_{v_{0,i}\in Q_0} (-1)^{x_i} \Bigg\rvert^2 = \begin{cases}
                                            1, \text{~if $x$ is balanced}\\
                                            0, \text{~if $x$ is constant}\\
                                        \end{cases}
    \end{equation*}
\end{proof}

\begin{theorem}
\label{thm:gqbp-parity}
    $Parity_n \in \gqbpcl{2}{n/2}$.
\end{theorem}
\begin{proof}
     Observe that any classical branching program that solves the problem with probability 1 must query all the bits in some path and so must have length $n$. To reduce the length, our \gqbp uses superposition to query each half separately, ``compute'' their parity in each branch, and then use a simple decision graph to output the xor of those parities. The \gqbp is illustrated in Appendix (see Figure~\ref{fig:gqbp-parity}).~\footnote{It is tempting to try this idea to reduce the length even further, e.g., to $n/4$ by creating a superposition of 4 branches. That approach would get stuck at the last stage of combining all the partial parities. In fact, we show later that it is not possible to reduce the length beyond $n/2$~(\ref{lemma:length-lower-bound}).}

    Without loss of generality, we assume that $n$ is even.
    To construct a \gqbp for $Parity_n$, set $Q = \big\{v_{i,a} : i\in \{0,1,\cdots, n/2\}, a\in \{0,1\}\big\}$, $\ket{v_0} = \frac{1}{\sqrt{2}}(\ket{v_{0,0}}+\ket{v_{0,1}})$, $E = \big\{(v_{i,a}, v_{i+1,b}) : i\in \{0,1,\cdots, n/2\}, a\in \{0,1\}\big\}$ and $F = \{v_{n/2, 1}\}$.
    Let $l=n/2$.
    As for the transition function $\delta$, for the first $l-1$ layers, i.e., $i\in \{0, 1, \cdots, l-2\}$ define
    \begin{equation*}
        \delta(v_{i,j}, a, v_{p,q}) = \begin{cases}
                                          1 \text{,~if~} p=i+1, j=q, a=0\\
                                          -1 \text{,~if~} p=i+1, j=q, a=1\\
                                          0 \text{,~else}
                                      \end{cases}
    \end{equation*}
    This gives that $U_i^{O_x} \ket{v_{i-1,a}} = (-1)^{L(v_{i-1,a})}\ket{v_{i,a}}$ for all $a\in \{0,1\}$.
    For the nodes $v_{l-1, 0}$ and $v_{l-1, 1}$, define
    \begin{equation*}
        \delta(v_{l-1,0}, 0, v_{l,0}) = \delta(v_{l-1,0}, 0, v_{l,1}) = \delta(v_{l-1,1}, 0, v_{l,0}) = \delta(v_{l-1,1}, 1, v_{l,1}) = 1/\sqrt{2}    
    \end{equation*}
    and
    \begin{equation*}
        \delta(v_{l-1,0}, 1, v_{l,0}) = \delta(v_{l-1,0}, 1, v_{l,1}) = \delta(v_{l-1,1}, 1, v_{l,0}) = \delta(v_{l-1,1}, 0, v_{l,1}) = -1/\sqrt{2}.
    \end{equation*}
    So, we have 
    \begin{align*}
        &U_{l-1}^{O_x}\ket{v_{l-1,0}} = \frac{(-1)^{L(v_{l-1,0})}}{\sqrt{2}}\Big(\ket{v_{l,0}}+\ket{v_{l,1}}\Big)\\
        &\text{and}\\
        &U_{l-1}^{O_x}\ket{v_{l-1,1}} = \frac{(-1)^{L(v_{l-1,1})}}{\sqrt{2}}\Big(\ket{v_{l,0}}-\ket{v_{l,1}}\Big).
    \end{align*}
    
    Next, define the function $L$ as $L(v_{i,j}) = 2*i+j$. Then, starting from the state $\ket{v_0}$, the state evolves through the first $l-1$ layers as follows,
    \begin{align*}
        \ket{v_0} &= \frac{1}{\sqrt{2}}(\ket{v_{0,0}} + \ket{v_{0,1}})\\
        &\xrightarrow{U_1^{O_x}} \frac{1}{\sqrt{2}}((-1)^{x_0}\ket{v_{1,0}} + (-1)^{x_1}\ket{v_{1,1}})\\
        &\xrightarrow{U_2^{O_x}} \frac{1}{\sqrt{2}}((-1)^{x_0\oplus x_2}\ket{v_{2,0}} + (-1)^{x_1\oplus x_3}\ket{v_{2,1}})\\
        &\cdots \\
        & \xrightarrow{U_i^{O_x}} \frac{1}{\sqrt{2}}((-1)^{x_0\oplus x_2\oplus \cdots \oplus x_{2i-2}}\ket{v_{i,0}} + (-1)^{x_1\oplus x_3\oplus \cdots \oplus x_{2i-1}}\ket{v_{i,1}})\\
        &\cdots\\
        &\xrightarrow{U_{l-1}^{O_x}}\frac{1}{\sqrt{2}}((-1)^{x_0\oplus x_2\oplus \cdots \oplus x_{n-4}}\ket{v_{l-1,0}} + (-1)^{x_1\oplus x_3\oplus \cdots \oplus x_{n-3}}\ket{v_{l-1,1}}) = \ket{\psi_{l-1}} \text{~(say)~}\\
    \end{align*}
    For the last transition, the state evolves as
    \begin{align*}
        \ket{\psi_{l-1}} &= \frac{1}{\sqrt{2}}((-1)^{x_0\oplus x_2\oplus \cdots \oplus x_{n-4}}\ket{v_{l-1,0}} + (-1)^{x_1\oplus x_3\oplus \cdots \oplus x_{n-3}}\ket{v_{l-1,1}})\\
        &\xrightarrow{U_{l}^{O_x}} \frac{1}{\sqrt{2}}\Bigg[(-1)^{x_0\oplus x_2\oplus \cdots \oplus x_{n-4}}\Big[\frac{(-1)^{x_{n-2}}}{\sqrt{2}}\Big(\ket{v_{l,0}+\ket{v_{l,1}}}\Big)\Big] \\
        &\hspace{1.5cm}+ (-1)^{x_1\oplus x_3\oplus \cdots \oplus x_{n-3}}\Big[\frac{(-1)^{x_{n-1}}}{\sqrt{2}}\Big(\ket{v_{l,0}}-\ket{v_{l,1}}\Big)\Big]\Bigg]\\
        &= \frac{1}{2}\Bigg[ \Bigg((-1)^{x_0\oplus x_2\oplus \cdots \oplus x_{n-2}}+(-1)^{x_1\oplus x_3\oplus \cdots \oplus x_{n-1}}\Bigg) \ket{v_{l,0}} \\
        &\hspace{1.5cm} +\Bigg((-1)^{x_0\oplus x_2\oplus \cdots \oplus x_{n-2}}-(-1)^{x_1\oplus x_3\oplus \cdots \oplus x_{n-1}}\Bigg) \ket{v_{l,1}}\Bigg]
    \end{align*}
    So, if $x_{0}\oplus x_{2}\oplus \cdots x_{n-2} = x_{1}\oplus x_{3}\oplus \cdots x_{n-1}$ or alternatively, $Parity_n(x) = 0$, then $\ket{\psi_l} = \ket{v_{l,0}}$ and the input will be rejected with probability $1$ post measurement.
    Else if $x_{0}\oplus x_{2}\oplus \cdots x_{n-2} \neq x_{1}\oplus x_{3}\oplus \cdots x_{n-1}$, i.e, if $Parity_n(x) = 1$, then $\ket{\psi_l} = \ket{v_{l,1}}$.
    In this case, the input will be accepted with probability $1$ post measurement.
    
\end{proof}

It is noteworthy that any query circuit requires at least $\log(n)$ qubits to query the oracle for an index.
Then as a derivative of Corollary~\ref{corr:circ-to-gqbp-optimized}, one would be led to think that any \gqbp would require width at least $n$.
However, in contradiction to that, Theorem~\ref{thm:gqbp-parity} gives us a width-2 \gqbp that computes the $Parity_n$ function exactly.
This is a classic example that shows that space lower bounds on query circuits does not hold for the GQBPs.
On the contrary, later in section~\ref{sec:rel-gqbp-circ}, we will show that asymptotic query lower bounds of query circuits also hold for the GQBPs and that of GQBPs hold for query circuits.

\begin{lemma}
    \label{lemma:maj3-gqbp}
    $Maj_3 \in \gqbpcl{4}{2}$
\end{lemma}
\begin{proof}
    The idea behind the construction is first to compare the first two bits. If they are the same, their value is the majority. Otherwise, the value of the third bit turns out as the majority. The state diagram is illustrated in Figure~\ref{fig:gqbp-maj3}.

    Consider a 3-levelled \gqbp \scrg$=(Q,E,\ket{v_0},L,\delta,F)$ for $Maj_3$ where the set of nodes $Q = \{v_{0,0}, v_{0,1}, v_{1,0}, v_{1,1}, v_{2,0}, v_{2,1}, v_{2,2}, v_{2,3}\}$, $\ket{v_0} = \frac{1}{\sqrt{2}}(v_{0,0} + v_{0,1})$, the function $L$ is such that $L(v_{0,0})=L(v_{1,0})=x_1, L(v_{0,1})=x_2$ and $L(v_{1,1}=x_3)$, and $F = \{\ket{v_{2,1}}, \ket{v_{2,3}}\}$.
    Define the transition function $\delta$ as 
    \begin{align*}
        &\delta(v_{0,0}, 0, v_{1,0}) = \delta(v_{0,0}, 0, v_{1,1}) = \delta(v_{0,1}, 0, v_{1,0}) = \delta(v_{0,1}, 1, v_{1,1}) = 1/\sqrt{2}\\
        &\delta(v_{0,0}, 1, v_{1,0}) = \delta(v_{0,0}, 1, v_{1,1}) = \delta(v_{0,1}, 1, v_{1,0}) = \delta(v_{0,1}, 0, v_{1,1}) = -1/\sqrt{2}\\
        &\text{and}\\
        &\delta(v_{1,0}, 0, v_{2,0}) = \delta(v_{1,0}, 1, v_{2,1}) = \delta(v_{1,1}, 0, v_{2,2}) = \delta(v_{1,1}, 0, v_{2,3}) = 1.
    \end{align*}

    From this transition function, we obtain the transition unitary $U_1^{O_x}$ as 
    \begin{align*}
        &U_{1}^{O_x}\ket{v_{0,0}} = \frac{(-1)^{L(v_{0,0})}}{\sqrt{2}}\Big(\ket{v_{1,0}}+\ket{v_{1,1}}\Big) = \frac{(-1)^{x_0}}{\sqrt{2}}\Big(\ket{v_{1,0}}+\ket{v_{1,1}}\Big)\\
        &\text{and}\\
        &U_{1}^{O_x}\ket{v_{0,1}} = \frac{(-1)^{L(v_{0,1})}}{\sqrt{2}}\Big(\ket{v_{1,0}}-\ket{v_{1,1}}\Big) = \frac{(-1)^{x_1}}{\sqrt{2}}\Big(\ket{v_{1,0}}-\ket{v_{1,1}}\Big).
    \end{align*}

    Now, one step of the evolution of the program can be given as
    \begin{align*}
        \ket{v_0} &= \frac{1}{\sqrt{2}}(\ket{v_{0,0}}+\ket{v_{0,1}})\\
        &\xrightarrow{U_1^{O_x}} \frac{1}{\sqrt{2}}\Bigg(\frac{(-1)^{x_0}}{\sqrt{2}}\Big(\ket{v_{1,0}+\ket{v_{1,1}}}\Big) + \frac{(-1)^{x_1)}}{\sqrt{2}}\Big(\ket{v_{1,0}}-\ket{v_{1,1}}\Big)\Bigg)\\
        &= \frac{1}{2}\Bigg(\Big((-1)^{x_0} + (-1)^{x_1}\Big)\ket{v_{1,0}} + \Big((-1)^{x_0} - (-1)^{x_1}\Big)\ket{v_{1,1}}\Bigg)\\
        &=\begin{cases}
            \ket{v_{1,0}}, \text{~~if $x_0=x_1$}\\
            \ket{v_{1,1}}, \text{~~if $x_0\neq x_1$}
        \end{cases}\\
        &\xrightarrow{U_2^{O_x}}\begin{cases}
                                    \ket{v_{2,0}}, \text{~~if $x_0=x_1$ and $x_0=0$}\\
                                    \ket{v_{2,1}}, \text{~~if $x_0=x_1$ and $x_0=1$}\\
                                    \ket{v_{2,2}}, \text{~~if $x_0\neq x_1$ and $x_2=0$}\\
                                    \ket{v_{2,3}}, \text{~~if $x_0\neq x_1$ and $x_2=1$}\\
                                \end{cases}
    \end{align*}
    We have $F = \{\ket{v_{2,1}}, \ket{v_{2,3}}\}$. This implies that if the given input is such that $x_0=x_1$ and $x_0=1$ or $x_0\neq x_1$ and $x_2=1$, then the \gqbp accepts the input.
    Now, see that if $x_0=x_1$, clearly, the value of $x_0$ is the majority. However, if $x_0\neq x_1$, then the value of $x_2$ is the majority.
    Putting them together, we have that the \gqbp \scrg computes the $Maj_3$ function exactly.
\end{proof}

On careful observation, one can notice that the amplitudes of the basis states of the state of a \gqbp after the evolution is a polynomial of the input $x$ since, at each node, the outgoing edges are decided based on the value of the query made at that node.
Moreover, each step of the evolution (applying one $U_i^{O_x}$) increases the degree of the polynomial corresponding to the amplitudes by at most $1$.
So, clearly, the amplitude of any node of the final state of an $l$-length \gqbp can be given by a polynomial of degree of at most $l$.
For any \gqbp,  the probability of accepting a given input is the sum of the probabilities of obtaining the nodes in $F$ on measuring the final state.
Consequently, we have the probability that an $l$-length \gqbp accepts an input $x$ is a polynomial of degree at most $2l$.
This observation leads us to the following theorem.
\begin{theorem}
    Let $f$ be a Boolean function. Let $deg(f)$ and $\widetilde{deg}(f)$ be the exact degree and approximate degree of $f$. Then,
    \begin{enumerate}
        \item Any \gqbp that exactly computes $f$ should have length at least $deg(f)/2$.
        \item Any \gqbp that approximates $f$ should have length at least $\widetilde{deg}(f)/2$.
    \end{enumerate}
\end{theorem}

As a result, we obtain the following lower bound that shows that our constructions above are tight with respect to length.

\begin{theorem}
\label{lemma:length-lower-bound}
    $Maj_3 \notin \gqbpcl{w}{1}$ and $Parity_n \not\in \gqbpcl{w}{t}$ for any $w\in \mathbb{N}$ and any $t<\tfrac{n}{2}$.
\end{theorem}

\section{Relation Between \gqbp~and other variants}
\label{sec:rel-gqbp-qbp}



\begin{theorem}
\label{thm:aqbp-gqbp-tight-inclusion}
    $\aqbpcl{w}{l} \subsetneq \gqbpcl{w}{l}$
    \label{thm:aqbp-subsetneq-gqbp}
\end{theorem}
\begin{proof}
    First, we show that any $(w,l)$-\aqbp is also a $(w,l)$-\gqbp. Let \scra$=\langle T, \ket{\psi_0}, F_a\rangle$ be a $(w,l)$-\aqbp where $T = \Big(\langle j_i ,U_i(0), U_i(1) \rangle\Big)_{i=1}^l$.
    Construct a $(w,l)$-\gqbp \scrg=$(Q, E, v_0, L, \delta, F_g)$ as follows.
    Set $Q = \{v_{(s,t)} : s\in \{0, 1, \cdots, l\}, t\in \{0, 1, \cdots, d-1\}\}$ and $F_g = \{v_{(l,t)} : \ket{t}\in F_a\}$.
    Let $v_0 = v_{(0,0)}$.
    Now, define a function $L:Q\setminus F\longrightarrow \{1, 2 \cdots, n\}$ as $L(v_{(s,t)}) = j_s$ for all $t$.
    Define the transition function $\delta: Q \times \{0,1\} \times Q \longrightarrow \mathbb{C}$ as 
    \begin{equation*}
        \delta (v_{s,t}, a, v_{p,q}) 
        = \begin{cases}
              0 \text{ if } p\neq s+1\\
              U_s(a)[q,t] \text{ if } p= s+1\\
          \end{cases}
    \end{equation*}
    It is easy to see that $\delta$ is quantumly well-behaved (to prove).
    

    To show that \scrg simulates \scra exactly, it suffices to show that that the unitary $U_i^{O_x} = U_i(j_i)$ for all $i\in [l]$ since $\ket{v_{0,0}} = \ket{v_0}$ and $F_g = F$.
    Fix an input $x$. Let $i\in [l]$. Then from definition of \gqbp, we have that for any $v\in Q_i$
    \begin{equation*}
        U_i^{O_x} \ket{v} = \sum_{v'\in Q_{i+1}}\delta(v,L(v),w)\ket{v'}
    \end{equation*}
    We know that $L(v) = j_i$ for any $v\in Q_i$.
    So, we have 
    \begin{equation*}
        U_i^{O_x} = \sum_{v\in Q_i} \Bigg[\sum_{v'\in Q_{i+1}}\delta(v,L(v),v')\ket{v'}\Bigg] \bra{v} = \sum_{v\in Q_i} \sum_{v'\in Q_{i+1}}U_i(j_i)[v',v]\ket{v'} \bra{v} = U_i(j_i).
    \end{equation*}
    Hence, \scrg simulates \scra exactly.
    This shows that any arbitrary $(w,l)$-\aqbp can be simulated by an equivalent $(w,l)$-\gqbp.

    To show the tight inclusion, observe that an \aqbp requires length at least $n$ to query all the bits of an $n$-length input. Thus, $Parity_n \not\in \aqbpcl{w}{t}$ for any $w$ and any $t < n$; however, we showed in the earlier section that $Parity_n \in \gqbpcl{2}{n/2}$.
    
\end{proof}

Next, we turn to \nqbp. Observe that a \gqbp can be written as an \nqbp if we set $Q_{acc}=F$, $Q_{rej}$ as the states not in $F$ in the last layer of the \gqbp, and $Q_{non}$ as the rest of the states. Since the evolution of a \gqbp happens in layers, in none of the evolutions but the last one will any state in $\ket{E_{acc}}$ and $\ket{E_{rej}}$ be observed, and hence the evolution would continue uninterrupted. And since there is no non-$E_{non}$ node in the last layer, the evolution would stop after that. Thus, a size-$s$ length-$l$ \gqbp is also a size-$s$ length-$l$ \nqbp.

Next, we prove the reverse direction.

\begin{theorem}
    Any size $s$ length $l$ \nqbp~can be simulated by a width $s$ length $l$ \gqbp.
    \label{thm:nqbp-equals-gqbp}
\end{theorem}
\begin{proof}
    Let $\mathcal{N}$ be an \nqbp of size $s$ and length $l$. Now, make $l$ many copies of $\mathcal{N}$.
    For any edge between two nodes $v, v'$ in $\mathcal{N}$, remove that edge and add an edge between node $v$ of copy $i$ and node $v'$ of copy $i+1$ for all $i\in [l-1]$.
    Clearly, this program is levelled since at $i^{th}$ step, the state of the program is in some superposition of the nodes of copy $i$.
    Since, the size of $\mathcal{N}$ is $s$, the space of the superposition is $\mathcal{H}(s)$.
\end{proof}

\section{Relation Between GQBPs and Quantum Query Circuits}
\label{sec:rel-gqbp-circ}

To the best of our knowledge, there does not exist any study of the relation between quantum circuits and quantum branching programs.
This section presents complete reductions from the GQBPs to the quantum query circuits and vice-versa.
As an add-on, we also present reductions from the AQBPs to the quantum circuits with QRAM and standard oracle access to inputs.

\begin{theorem}
\label{thm: qram-circuit-to-aqbp}
    For any $q$-qubit, $t$-query quantum circuit $C$ with QRAM query model, there exists an $\big(2^q, t\big)$-\aqbp~that simulates $C$.
\end{theorem}
\begin{proof}
    Let $C$ be a $q$-qubit $t$-query quantum circuit with a QRAM query model.
    Let $C$ be given as $C = \Tilde{O}^{p_t}\cdots \Tilde{O}^{p_1}$.
    where QRAM oracle $\Tilde{O}^{p_k}$ can be defined as
    \begin{equation*}
        \Tilde{O}^{p_k} = \mathbb{I}^{p_k-1} \otimes \ket{1}\bra{1} \otimes \mathbb{I}^{n-p_k} \otimes \Tilde{U}^{(0)}_{k} + \mathbb{I}^{p_k-1} \otimes \ket{1}\bra{1} \otimes \mathbb{I}^{n-p_k} \otimes \Tilde{U}^{(1)}_{k}
    \end{equation*}
    Without loss of generality, let the initial state be $\ket{0}$.
    Let $F_c$ be the set of basis states which, on obtaining post-measurement the given input is accepted.

    Construct an \aqbp \scra$=\langle T, \ket{\psi_0}, F_A \rangle$ as follows.
    Set $\ket{\psi_0} = \ket{0}$ and $F_A=F_c$.
    Define $T = (\langle j_i, U_i(0), U_i(1) \rangle)_{i=1}^t$ where for any $i\in \{1,\cdots, t\}$, we set $j_i = p_i$, $U_i(0) = \Tilde{U}_i^{(0)}$ and $U_i(1) = \Tilde{U}_i^{(1)}$.
    We can see that $U_i(0), U_i(1)$ unitaries act on the Hilbert space $\mathcal{2^q}$.
    Clearly, the final state of \scra before the measurement can be obtained as
    \begin{equation*}
        \ket{\psi_f} = U_t(j_t)\cdots U_1(j_1)\ket{\psi_0} = \Tilde{U}_t^{(j_t)}\cdots \Tilde{U}_1^{(j_1)}\ket{0} = C\ket{0}.
    \end{equation*}
    Since we have $F_A = F_c$, the probability of accepting an input $x$ by \scra is
    \begin{equation*}
        Pr_{\mathcal{A}}[accept | x] = \sum_{\ket{i}\in F_A} \Big\lvert \bra{i}\ket{\psi_f} \Big\rvert^2 = \sum_{\ket{i}\in F_c} \Big\lvert \bra{i}C\ket{0} \Big\rvert^2 = Pr_{C}[accept | x].
    \end{equation*}
    This gives that the $(2^q, t)$-\aqbp \scra exactly simulates the circuit $C$.
\end{proof}

\begin{corollary}
\label{cor: qram-circuit-to-gqbp}
    For any $q$-qubit, $t$-query quantum circuit $C$ with QRAM query model, there exists an $\big(2^q, t\big)$-\gqbp~that simulates $C$.
\end{corollary}

We show the following reductions from an \aqbp~to a quantum circuit.
\begin{theorem}
\label{thm: aqbp-to-circuit}
\begin{enumerate}
    \item Any $(d,l)$-\aqbp~ can be simulated using a circuit with $\log d$ qubits and $l$ gates that have a control on the input in the QRAM model when the input is stored in the QRAM.
    \item Any $(d,l)$-\aqbp~ can be simulated using a circuit with $\log n + \log d + 1$ qubits and $2\cdot l$ queries to the oracle when the input is accessible only using an oracle.
\end{enumerate}    
\end{theorem}

The proof of Theorem~\ref{thm: aqbp-to-circuit} is presented in Appendix~\ref{appendix:aqbp-to-circ}.
Although the reduction from an \aqbp to a query circuit looks trivial, the same is not true for a reduction from a \gqbp to a query circuit.
Moreover, a generic reduction from a \gqbp to a query circuit under the QRAM model is impossible.

We now show that any quantum query circuit can be simulated by a \gqbp exactly.

\begin{theorem}
    \label{thm:circ-to-gqbp}
    For any $q$-qubit $t$-query quantum circuit $C$ that has access to the input through a standard oracle, there exists an $\big(2^q, 2t+1\big)$-\gqbp~that simulates $C$.
\end{theorem}
\begin{proof}
    On an outline, to simulate $C$, we construct a \gqbp such that all the odd transitions correspond to the unitaries and the even transitions correspond to the oracle calls.
    For this proof we assume that the query oracle used is a phase oracle, i.e., the oracle $O_x$ acts as $O_x\ket{i} = (-1)^{x_i}\ket{i}$.
    Let the circuit $C$ be given as $C = \Tilde{U}_tO_x\Tilde{U}_{l-1}\cdots \Tilde{U}_1O_x\Tilde{U}_0$ and let $F_c$ be the set of basis states which on obtaining post measurement the given input is accepted.
    Fix $l=2t+1$ and $w = 2^q$.
    Now, define a \gqbp \scrg$=(Q,E,v_0,L,\delta,F)$. Set $Q = \{v_{i,j}: i\in \{0,1,\cdots,l\}, j\in \{0,1,\cdots,w-1\} \}$, $E = \{(v_{i,j}, v_{i+1,k}) : i\in \{0,1,\cdots, l\} \text{~and~} j,k\in \{0,1,\cdots, w-1\}\}$ and $\ket{v_0} = \ket{v_{0,0}}$
    Also, fix $F = \{v_{l,j} : \ket{j}\in F_c\}$.
    Define the transition function $\delta$ as follows:
    \begin{equation*}
        \delta(v_{i,j}, a, v_{p,q}) = \begin{cases}
                                            \Tilde{U}_{i/2}[q,j]~~\forall~a\in\{0,1\}\text{,~~if $i$ is even and $p=i+1$}\\
                                            1 \text{,~~if $j=q$, $a=0$, $i$ is odd and $p=i+1$}\\
                                            -1 \text{,~~if $j=q$, $a=1$, $i$ is odd and $p=i+1$}\\
                                            0 \text{,~~else}
                                        \end{cases}
    \end{equation*}
    It is clear that for any transition between the $2i^{th}$ and $(2i+1)^{th}$ layer, the transition is query independent and the unitary $U_{2i}^{O_x} = \Tilde{U}_{i}$.
    This simulates the query-independent unitaries of the circuit $C$.
    Meanwhile, the unitary $U_{2i+1}^{O_x}$ is query dependent and is a diagonal matrix where the $j^{th}$ entry takes the value $+1$ if $x_{L(v_{2i+1,j})}=1$ and $-1$ otherwise, i.e., the unitary acts as
    \begin{equation*}
        U_{2i+1}^{O_x} \ket{v_{2i+1,j}} = (-1)^{x_{L(v_{2i+1,j})}}\ket{v_{2i+1,j}}.
    \end{equation*}
    We can see that the unitary $U_{2i+1}^{O_x}$ exactly simulates a query to the oracle $O_x$.

    So, the final state after the evolution of the branching program can be given as
    \begin{align*}
        \ket{\psi_f} = U_{l}^{O_x}\cdots U_1^{O_x}\ket{v_0} = \Tilde{U}_{l/2}O_x\Tilde{U}_{l/2-1}\cdots \Tilde{U}_1 O_x \Tilde{U_0}\ket{v_0}
    \end{align*} 
    Which is equivalent to the final state of the circuit $C$.
    Finally, on measuring $\psi_f$ and obtaining an output $v_{l,j}$, an input $x$ is accepted iff $v_{l,j}\in F$ or equivalently iff $\ket{j}\in F_c$.
    So, the probability of accepting an input $x$ by \scrg is
    \begin{equation*}
        Pr[accept] = \sum_{v_{l,j}\in F} \Big\lvert \bra{v_{l,j}}\ket{\psi_f}\Big\rvert^2 = \sum_{j\in F_c} \Big\lvert \bra{j}C\ket{0}\Big\rvert^2
    \end{equation*}
    This shows that the $(2^q,2t+1)$-\gqbp \scrg simulates the query circuit $C$ exactly.
\end{proof}

In the above reduction note that the unitaries $U_{2i}^{O_x}$ are query independent.
We call the nodes that evolve in a query-independent way as `\textbf{dummy}' nodes.
In the next lemma, we show that any dummy node in a \gqbp is removable without any overhead but with a possible reduction in width and length.

\begin{lemma}
\label{lemma:dummy-nodes}
    Any dummy node in a \gqbp is removable without any width or length overhead.
\end{lemma}
\begin{proof}
    Consider a \gqbp \scrg with 3 levels. Without loss of generality, let all levels contain $k$ nodes, and the second level only contains dummy nodes.
    Let $Q_i = \{v_{i,j} : j\in \{0,1,\cdots,k-1\}\}$ for $i\in \{0,2\}$ and $Q_2 = \{d_0, d_1 \cdots, d_{k-1}\}$.
    Fix an arbitrary input $x$. Per the definition, the transition matrix $U_1^{O_x}$ from level $1$ to level $2$ must be unitary.
    Similar is the case for the transition function $U_2^{O_x}$ for evolution from level $2$ to level $3$.

    Let $U_c^{O_x} = U_2^{O_x}\cdot U_1^{O_x}$. We show that the \gqbp \scrg can be replaced by another \gqbp \scrg', which has only two levels.
    Consider two nodes $v_{0,i}$ and $v_{3,j}$ of \scrg. Then, on considering all the paths from $v_{0,i}$ to $v_{2,j}$, we can get the combined transition amplitude as
    \begin{equation*}
        \sum_{l=0}^{k-1}\delta(d_l, b, v_{2,j})\delta(v_{0,i}, a, d_l) = \sum_{l=0}^{k-1}U_2^{O_x}[j,l]U_1^{O_x}[l,i] = U_c^{O_x}[j,i].
    \end{equation*}
    See that this amplitude is independent of $b$ and dependent only on $a$ since $d_l$s are dummy nodes.
    So, it is possible to construct a transition function that is defined as 
    $\delta'(v_{0,i}, a, v_{2,j}) = \sum_{l=0}^{k-1}\delta(d_l, b, v_{2,j})\delta(v_{0,i}, a, d_l)$ that mimics the transition from the node $v_{0,i}$ to $v_{2,j}$ in \scrg.
    Since this is true for any two nodes $v_{0,i}\in Q_0$ and $v_{1,i}\in Q_2$, we can replace the \gqbp \scrg with a \gqbp \scrg that has only two levels with $Q_0$ and $Q_2$ for nodes and with transition function $\delta'$ for transitions between nodes in $Q_0$ and $Q_2$.
    
\end{proof}

We can now use Lemma~\ref{lemma:dummy-nodes} to tighten the result in Theorem~\ref{thm:circ-to-gqbp} to obtain the following corollary.

\begin{corollary}
    \label{corr:circ-to-gqbp-optimized}
    For any $q$-qubit $t$-query quantum circuit $C$, there exists an $\big(2^q, t\big)$-\gqbp~that simulates $C$.
\end{corollary}

From this result, one would hope that any $(w,l)-$\gqbp would be reducible to a $\log w$-width $l$-query circuit.
However, the \gqbp we presented for the $Parity_n$ function serves a a contradiction to this.
To see this, note that any query circuit would require at least $\log n$ qubits (that correspond to the indices) that would be used to query the oracle.
So, any query circuit computing the $Parity_n$ function exactly will need at least $\log n$ qubits and $n/2$ queries.
The lower bound on the number of queries comes from the facts that $Q_E(f) \ge deg(f)/2$ and that $deg(Parity_n)=n$.
Now, if the above assumption were true, then clearly, we get a lower bound on $w$ as $w \ge n$.
However, the \gqbp presented in Section~\ref{sec:upper-bounds} for the $Parity_n$ function has length $l=n/2$ but width $2$ that contradicts this lower bound.

We now present the following theorem that characterizes a reduction from a \gqbp to a query circuit.

\begin{theorem}
\label{thm:gabp-to-circ}
    For any $(w,l)$-\gqbp \scrg, there exists an $(l\log w + \log n + 1)$-qubit $2l$-query circuit $C$ with access to input through an oracle $O_x$ that simulates \scrg exactly.
\end{theorem}

The proof of this theorem is presented in Appendix~\ref{appendix:gqbp-to-circ}.

\section{Conclusion and Open Directions}
\label{sec:conclusion}

In this work we present a quantum variant of a decision graph~\cite{masek}, commonly known as a branching program, that mimics the evolution of a probabilistic (classical) branching program and further enables it to do so in a superposition. We are primarily concerned about the relationship of quantum query circuits and we prove various results explaining why and how they are equivalent (or, not).

There are several interesting questions that wait to be answered.

What would be the relation between (non-query) quantum circuits or quantum TMs and \gqbp, especially with respect to the number of ancill\ae\ qubits of the former and the space complexity of the latter? For example, it is known that polynomial-size classical branching programs are equivalent to non-uniform logspace classical TMs and poly-size Boolean circuits. Many results about classical branching programs rely on counting arguments and that certainly does not readily work in the quantum case.

We showed that an \aqbp is strictly weaker than a \gqbp by essentially drawing attention to the fact that the former requires length $n$ to be able to query {\em every input bit} which the latter does not. This actually stems from the structural requirement of \aqbp to query the same variable in a level. Thus, to really compare the strength of the two models, one should consider programs of length $\Omega(n)$, and therefore, of bounded sub-exponential width. We conjecture that a \gqbp of length $t=\Omega(n)$ and width $w=O(poly(n))$ is strictly stronger than an \aqbp of similar length and width.

It is often assumed that a quantum model can do everything that a classical model can do, and even more. However, that does not appear to hold here. Specifically, it is not clear how to simulate a deterministic classical branching program of width $w$ and length $l$ using a \gqbp of width $O(w)$ and length $O(l)$. The difficulty lies in the requirement that the transition function of a \gqbp has to be quantumly well-behaved.

The above question is trivially resolved if we instead consider simulating a {\em Permutation Branching Programs} (\pbp). A \pbp is a classical levelled branching program that queries the same input bit in a level, and depending upon its value, performs a permutation on the nodes in that level (observe that a permutation matrix is also unitary). An \aqbp is a quantum version of a \pbp. Barrington's seminal result about $NC^1$ showed how to simulate any problem in that class using a width-5 (polynomial length) \pbp~\cite{barrington1989150}. Improving this result, Ablayebv et al.~\cite{ablayev2002quantum} proved only width 2 is necessary to simulate any $NC^1$ problem using an \aqbp. Based on our understanding of \gqbp, we conjecture that a bounded-width \gqbp exists for any $QNC^1$ problem.


\bibliography{refs}

\newpage

\appendix
\section{A \gqbp for the Deutsch-Jozsa problem}
\label{appendix:parity-gqbp}

We provide here a pictorial representation of the \gqbp for the $Parity_n$ problem.

\begin{figure}[h!]
\centering
        \begin{tikzpicture}[node distance={30mm}, thick, main/.style = {draw, circle}, empty/.style = {circle}, rect/.style={draw, rectangle}] 
            \node[main] (0) {$x_0$}; 
            \node[main] (1) [right of=0] {$x_1$};
            \node[main] (2) [below of=0] {$x_2$};
            \node[main] (3) [below of=1] {$x_3$};
            \node[empty] (4) [below of=2] {$\vdots$};
            \node[empty] (5) [below of=3] {$\vdots$};
            \node[main] (6) [below of=4] {\scalebox{0.75}{$x_{n\text{-}2}$}};
            \node[main] (7) [below of=5] {\scalebox{0.75}{$x_{n\text{-}1}$}};
            \node[rect] (8) [below of=6] {$0$};
            \node[rect] (9) [below of=7] {$1$};
            \node[empty](10) [above of=0, yshift=-40pt] {};
            \node[empty](11) [above of=1, yshift=-40pt] {};

            \draw[->, dotted] (10) to node[left] {$\frac{1}{\sqrt{2}}$} (0);
            \draw[->, dotted] (11) to node[right] {$\frac{1}{\sqrt{2}}$} (1);
            
            \draw[->, green, dashed] (0) to [out=-135,in=135,looseness=1] node[left] {-1} (2);
            \draw[->, red] (0) to [out=-45, in=45,looseness=1] node[right] {1} (2);
            \draw[->, green, dashed] (1) to [out=-135,in=135,looseness=1] node[left] {-1} (3);
            \draw[->, red] (1) to [out=-45, in=45,looseness=1] node[right] {1} (3);
            \draw[->, green, dashed] (2) to [out=-135,in=135,looseness=1] node[left] {-1} (4);
            \draw[->, red] (2) to [out=-45, in=45,looseness=1] node[right] {1} (4);
            \draw[->, green, dashed] (3) to [out=-135,in=135,looseness=1] node[left] {-1} (5);
            \draw[->, red] (3) to [out=-45, in=45,looseness=1] node[right] {1} (5);
            \draw[->, green, dashed] (4) to [out=-135,in=135,looseness=1] node[left] {-1} (6);
            \draw[->, red] (4) to [out=-45, in=45,looseness=1] node[right] {1} (6);
            \draw[->, green, dashed] (5) to [out=-135,in=135,looseness=1] node[left] {-1} (7);
            \draw[->, red] (5) to [out=-45, in=45,looseness=1] node[right] {1} (7);
            \draw[->, green, dashed] (6) to [out=-135,in=135,looseness=1] node[left] {+} (8);
            \draw[->, red] (6) to node[right] {$-$} (8);
            \draw[->, green, dashed] (6) to [out=-60,in=170,looseness=0.75] node[below right=30pt] {+} (9);
            \draw[->, red] (6) to [out=-30,in=120,looseness=0.75] node[above] {$-$} (9);
            \draw[->, green, dashed] (7) to [out=200,in=45,looseness=0.75] node[below left=20pt] {+} (8);
            \draw[->, red] (7) to [out=230,in=20,looseness=0.75] node[below] {$-$} (8);
            \draw[->, green, dashed] (7) to [out=-45,in=45,looseness=1] node[right] {$-$} (9);
            \draw[->, red] (7) to node[right] {+} (9);

            \matrix [draw,below right] at (current bounding box.north east) {
                \draw[-latex,color=black, dotted](0,-0.3) -- (0.6,-0.3); & \node{Initial edge};\\
                \draw[-latex,color=green, dashed](0,-0.3) -- (0.6,-0.3); & \node{$0$-edge};\\
                \draw[-latex,color=red](0,-0.3) -- (0.6,-0.3); & \node{$1$-edge};\\
            };
            
        \end{tikzpicture} 
        \caption{A \gqbp for the $n$-bit Parity problem. $+$ and $-$ indicates transitions with amplitudes $\tfrac{1}{\sqrt{2}}$ and $-\tfrac{1}{\sqrt{2}}$, respectively.\label{fig:gqbp-parity}}
    \end{figure}
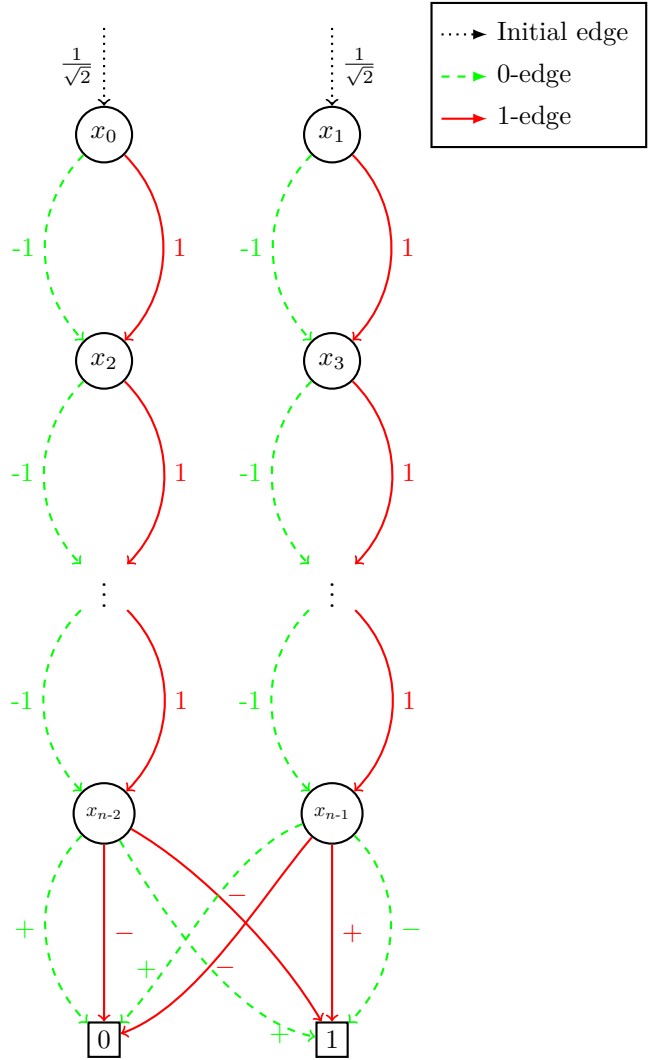

\section{Proof of Theorem 18}
\label{appendix:aqbp-to-circ}

We first proof the reduction from an \aqbp to a query circuit under the QRAM model.

\begin{proof}
    Let the input string be stored in QRAM.
    At any particular level of a $(d,l)$-\aqbp, the maximum number of nodes is $d$.
    At each level, denote each node by a basis state from $\ket{0}$ to $\ket{d-1}$.
    Then, the total number of qubits required to express the state of a $(d,l)$-\aqbp~is $\log(d)$ since after the $i^{th}$ step of evolution, the state would be in a superposition of nodes of level $i$ (we assume the source node is at level $0$.)
    
    Next, at iteration $i$, we perform a quantum transformation of the type $\langle j_i, U_i(0), U_i(1) \rangle$ that takes a superposition of nodes at level $i-1$ to a superposition of nodes at level $i$.
    Alternatively, for $a\in \{0,1\}$, $U_i(a)$ maps each node at level $i-1$ to a superposition of nodes at level $i$.
    Since we denote the nodes by computational basis states, let $\what{U}_i(a)$ be the unitary that corresponds to the unitary $U_i(a)$ in the computational basis state notation.

    Now, construct the circuit $C_P$ that simulates the $(d,l)$-\aqbp~$P$ as follows:
    \begin{enumerate}
        \item Initialize the circuit with $\log d$ qubits with the initial state as the state $\ket{0}$ corresponding to the source node $v_0$.
        \item For $i=1$ to $l$: \label{line: iteration_step_basic}
        \begin{enumerate}
            \item Apply $\what{U}_i(0)$.
            \item Controlled on the $j_{i}^{th}$ qubit in QRAM being $1$, apply $\what{U}^\dagger_i(0)\what{U}_i(1)$.
        \end{enumerate}
        \item Measure the final state $\ket{\psi_l}$ in the computational basis. If the measurement outcome corresponds to a node in $F$, then accept the input. Else, reject the input.
    \end{enumerate}

    We show the correctness of the circuit $C_P$ now.
    To prove this it suffices to show that the final state obtained just before measurement corresponds to the final state obtained in the \aqbp~P, i.e, if the circuit outputs the state
    \begin{equation*}
        \ket{\psi_l} = \what{U}_l(x_{j_l})\what{U}_{l-1}(x_{j_{i-1}})\cdots \what{U}_2(x_{j_2})\what{U}_1(x_{j_1})\ket{\psi_0}
    \end{equation*}
    The circuit starts in the state $\ket{0} = \ket{\psi_0}$ that corresponds to the node $v_0$.
    Next, at iteration $i$, in step~\ref{line: iteration_step_basic} we apply only $\what{U}_i(0)$ to $\ket{\psi_{i-1}}$ if $x_{j_i}$ qubit is $0$.
    Else, we apply $\what{U}_i(0)\what{U}^\dagger_i(0)\what{U}_i(1) = \what{U}_i(0)$ to $\ket{\psi_{i-1}}$ if $x_{j_i} = 1$.
    So, the final state $\ket{\psi_l}$ we obtain is of the form,
    \begin{equation*}
        \ket{\psi_l} = \what{U}_l(x_{j_l})\what{U}_{l-1}(x_{j_{i-1}})\cdots \what{U}_2(x_{j_2})\what{U}_1(x_{j_1})\ket{\psi_0}
    \end{equation*}
    as expected.
    Finally, we measure $\ket{\psi_l}$ to see if the measurement outcome lies in the set of basis states that correspond to the accepting set $F$.

    It is straightforward that the total number of qubits used in $C_P$ is $\log d$ and the total number of controlled gates used is at most $l$.
\end{proof}

Next, we prove the reduction from an \aqbp to a query circuit under the standard oracle model.

\begin{proof}
    We make the node to basis state mapping and $U_i(a)$ to $\what{U}_i(a)$ mappings for $a\in \{0,1\}$ as done for the QRAM setup.
    However, now since the query is made to an oracle $O_x$ that contains the input, at each iteration we also need an index that needs to be queried at that iteration.
    For this, we need an extra $\log n$ qubits and $1$ qubit to store the value of the query.
    Let $S_i$ denote the $\log n$ qubit unitary that performs the mapping $\ket{0} \longrightarrow \ket{j_i}$.
    Here we assume the oracle $O_x$ as the standard oracle that acts on a $\log n + 1$ qubits state as $O_x\ket{a}\ket{b} \longrightarrow \ket{a}\ket{b\oplus x_a}$.

    Construct the circuit $C_P$ corresponding to the $(d,l)$-\aqbp~P as follows.
    \begin{enumerate}
        \item Initialize the circuit with $3$ registers $R_1R_2R_3$ of sizes $\log n, \log d$ and $1$ respectively with initial states as $\ket{0}$ in all the registers.
        \item For $i=1$ to $l$:
        \begin{enumerate}
            \item Apply $S_i$ on $R_1$ to obtain the state $\ket{j_i}$.
            \item Query the oracle $O_x$ and store the query in $R_3$.
            \item Controlled on $R_3$ being in the state $\ket{a}$, apply $\what{U}_i(a)$ for $a\in \{0,1\}$.
            \item Query the oracle $O_x$ to reset $R_3$ to $\ket{0}$.
            \item Apply $S^\dagger_i$ on $R_1$ to reset $R_1$ to $\ket{0}$.
        \end{enumerate}
        \item Measure $R_2$ of the final state $\ket{\phi_l}$ in the computational basis. If the measurement outcome corresponds to a node in $F$, then accept the input. Else, reject the input.
    \end{enumerate}

    Now, we show the correctness of the circuit $C_P$. We first initialize the circuit with $3$ registers $R_1R_2R_3$, the index register, the node register and the ancilla register, all set to $\ket{0}$.
    At iteration $i$, we perform the following operations:
    \begin{align*}
        \ket{\phi_{i-1}} = \ket{0}\ket{\psi_{i-1}}\ket{0} & \xrightarrow{S_i\otimes \mathbb{I}\otimes \mathbb{I}} \ket{j_i}\ket{\psi_{i-1}}\ket{0}\\
        & \xrightarrow{O_x} \ket{j_i}\ket{\psi_{i-1}}\ket{x_{j_i}}\\
        & \xrightarrow{\mathbb{I}\otimes \big[\what{U}_i(0) \otimes \ket{0}\bra{0} + \what{U}_i(1) \otimes \ket{1}\bra{1}\big]} \ket{j_i}\big[\what{U}_i(x_{j_i})\ket{\psi_{i-1}}\big]\ket{x_{j_i}} = \ket{j_i}\ket{\psi_{i}}\ket{x_{j_i}}\\
        & \xrightarrow{O_x} \ket{j_i}\ket{\psi_{i}}\ket{0}\\
        & \xrightarrow{S_i^\dagger\otimes \mathbb{I}\otimes \mathbb{I}} \ket{0}\ket{\psi_{i}}\ket{0} = \ket{\phi_i}\\
    \end{align*}
    So, the state of the $R_2$ register after $i^{th}$ iteration is
    \begin{equation*}
        \ket{\psi_i} = \what{U}_i(x_{j_i})\ket{\psi_{i-1}}.
    \end{equation*}
    So, the final state of $R_2$ register would\textbf{} be of the form
    \begin{equation*}
        \ket{\psi_l} = \what{U}_l(x_{j_l})\what{U}_{l-1}(x_{j_{i-1}})\cdots \what{U}_2(x_{j_2})\what{U}_1(x_{j_1})\ket{\psi_0}
    \end{equation*}
    as required.
    We then measure this register and check if the output is a basis state corresponding to some node in $F$ to accept or reject the input $x$.
    
    Clearly, the number of queries made to the oracle $O_x$ is $2\cdot l$ and the total number of qubits used is $\log n + \log d + 1$.
    
\end{proof}

\section{Proof of Theorem 22}
\label{appendix:gqbp-to-circ}

\begin{theorem}[Restated Theorem 22]
    For any $(w,l)$-\gqbp \scrg, there exists an $(l\log w + \log n + 1)$-qubit $2l$-query circuit $C$ with access to input through an oracle $O_x$ that simulates \scrg exactly.
\end{theorem}

\begin{proof}
    Fix an arbitrary width $w$. We now prove the reduction by induction on the length of \gqbp.
    Let \scrg=$(Q, E, \ket{v_0}, L, \delta, F)$ be a $(w,l)$-\gqbp.
    For ease of use we we denote $\delta(v_{i,j}, a, v_{i+1,k})$ as $\beta_{j,k,a}^{(i+1)}$.
    Let $l=1$.

    From the definition, for the only step of the evolution of this program, we define the unitary $U_{1}^{O_x}$ as 
    \begin{equation*}
        U_{1}^{O_x}\ket{v} = \sum_{v'\in Q_1}\delta(v, O_x(L(v)), v')\ket{v'} = \sum_{v'\in Q_1}\beta_{v,v', L(v_i)}^{(1)}\ket{v'}.
    \end{equation*}
    Then, the final state we obtain before the measurement is given by
    \begin{align*}
        \ket{\psi_1} &= U_{1}^{O_x}\ket{v_0}\\
        &= \sum_{v_i\in Q_0}\alpha_i U_{1}^{O_x}\ket{v_i}\\
        &= \sum_{v_i\in Q_0}\alpha_i \Bigg[\sum_{v_j\in Q_1}\beta_{v_i,v_j,L(v_i)}^{(1)}\ket{v_j}\Bigg]\\
        &= \sum_{v_j\in Q_1} \Bigg[\sum_{v_i\in Q_0}\alpha_i\beta_{v_i,v_j,L(v_i)}^{(1)}\Bigg]\ket{v_j}\\
    \end{align*}

    So, the probability of obtaining one of the accept states is 
    \begin{equation*}
        Pr_{\mathcal{G}}[accept] = \sum_{v_j\in F} \Bigg\lvert\sum_{v_i\in Q_0}\alpha_i\beta_{v_i,v_j,L(v_i)}^{(1)}\Bigg\rvert^2
    \end{equation*}
    
    Construct a circuit $\mathcal{C}_\mathcal{G}$ with two registers $R_1, R_2$ each of size $\lceil w \rceil$ qubits, one register $R_3$ of size $\log n$ and one register $R_4$ of size $1$ qubit all initialized to $\ket{0}$.
    Let $\mathcal{C}_{\mathcal{G}}$ be provided with an access to the input $X$ through the oracle $O_X$ and let $\widehat{U}_L$ be the unitary that acts as $\widehat{U}_L \ket{j}\ket{0} = \ket{j}\ket{L(v_{0,j})}$.
    Since we know that initial state of \scrg, we can construct a unitary $U_{init}$ that act as $U_{init}\ket{0} = \sum_{i\in [w]}\alpha_i\ket{i}$ apriori.
    Then one step of \scrg can be simulated in $\mathcal{C}_\mathcal{G}$ as follows:
    \begin{align*}
        \ket{0}\ket{0}\ket{0}\ket{0} &\xrightarrow{U_{init}} \sum_{i\in [w]}\alpha_i\ket{i}\ket{0}\ket{0}\ket{0}\\
        &\xrightarrow{CX(R_1, R_2)} \sum_{i\in [w]} \alpha_i \ket{i}\ket{i}\ket{0}\ket{0}\\
        &\xrightarrow{\widehat{U}_L(R_1, R_3)} \sum_{i\in [w]} \alpha_i \ket{i}\ket{i}\ket{L(v_{0,i})}\ket{0}\\
        &\xrightarrow{O_X(R_3, R_4)} \sum_{i\in [w]} \alpha_i \ket{i}\ket{i}\ket{L(v_{0,j})}\ket{X_{L(v_{0,j})}}\\
        &\xrightarrow{C-U_{1}(R_1, R_2, R_4)} \sum_{i\in [w]} \alpha_i \ket{i}\Big(\sum_{j\in [w]} \beta_{i,j,X_i}^{(1)}\ket{j}\Big)\ket{L(v_{0,j})}\ket{X_{L(v_{0,j})}}\\
        &\xrightarrow{O_X(R_3, R_4)} \sum_{i\in [w]} \alpha_i \ket{i}\Big(\sum_{j\in [w]} \beta_{i,j,X_i}^{(1)}\ket{j}\Big)\ket{L(v_{0,j})}\ket{0}\\
        &\xrightarrow{\widehat{U}_L(R_1, R_3)} \sum_{i\in [w]} \alpha_i \ket{i}\Big(\sum_{j\in [w]} \beta_{i,j,X_i}^{(1)}\ket{j}\Big)\ket{0}\ket{0}\\
        &= \sum_{j\in [w]} \Big(\sum_{i\in [w]} \alpha_i \beta_{i,j,X_i}^{(1)}\ket{i}\Big)\ket{j}\ket{0}\ket{0} = \ket{\psi_f}\text{~(say)}\\
    \end{align*}

    Here, $C-U_{1}$ is defined as
    \begin{equation*}
        C-U_{1} = \sum_{i\in [w]}\ket{i}\bra{i} \otimes \Big[\sum_{j\in [w]} \beta_{i,j,0}^{(1)}\ket{j}\bra{i} \otimes \ket{0}\bra{0} + \sum_{j\in [w]} \beta_{i,j,1}^{(1)}\ket{j}\bra{i} \otimes \ket{1}\bra{1}\Big].
    \end{equation*}
    Note that the unitary $C-U_1$ is query independent.
    Next, let $\mathcal{M} = {P_0, P_1}$ be a projective measurement where $P_1 = \sum_{v_i\in F}\ket{i}\bra{i}$ and $P_0 = I-P_1$.
    Once we measure the second register using $\mathcal{M}$,the probability of accepting $X$ can be computed as
    \begin{equation*}
        Pr_{\mathcal{C_G}}[accept] = \sum_{v_j\in F} \Bigg\lvert \sum_{i\in [w]} \alpha_i \beta_{i,j,X_i}^{(1)}\ket{i}\Bigg\rvert^2 = \sum_{v_j\in F} \Bigg\lvert \sum_{i\in [w]} \alpha_i \beta_{i,j,X_i}^{(1)}\Bigg\rvert^2.
    \end{equation*}
    Note that this is exactly the probability of accepting the input $X$ in \scrg.
    This gives that the $\log(w)+\log(n)+1$-qubit $2$-query $\mathcal{C}_\mathcal{G}$ simulates the $(w,1)$-\gqbp \scrg exactly.

    Now, assume the result true for some $l=t$. Let \scrg'$=(Q',E',\ket{v'_0},L',\delta', F')$ be a $(w,t+1)$-\gqbp.
    Let the state at the end of $t$ layers be given as
    \begin{equation*}
        \ket{\psi_t} = \sum_{v_i\in Q_t}\gamma_{i}\ket{v_i}.
    \end{equation*}
    The transition unitary $U_{t}^{O_x}$ of the last step of the evolution of \scrg' can be defined as 
    \begin{equation*}
        U_{t}^{O_x}\ket{v} = \sum_{v'\in Q'_{t+1}}\delta'(v, O_x(L'(v)), v')\ket{v'} = \sum_{v'\in Q'_{t+1}}\beta_{v,v', L'(v_i)}^{(t)}\ket{v'}.
    \end{equation*}
    Then, the final state of the \gqbp would be
    \begin{equation*}
        \ket{\psi_{t+1}} = U_{t}^{O_x}\ket{\psi_t} = \sum_{v_i\in Q_{t}}\gamma_j \Bigg[\sum_{v_j\in Q_{t+1}}\beta_{v,v_j, L'(v_i)}^{(t)}\ket{v_j}\Bigg] = \sum_{v_j\in Q_{t+1}} \Bigg[\sum_{v_i\in Q_t}\gamma_i \beta_{v,v_j, L'(v_i)}^{(t)}\Bigg] \ket{v_j}.
    \end{equation*}
    So, the probability of accepting an input $X$ is 
    \begin{equation*}
        Pr_{\mathcal{G'}}[accept] = \sum_{v_j\in F'} \Bigg\lvert\sum_{v_i\in Q_t}\gamma_i\beta_{v_i,v_j,L(v_i)}^{(t)}\Bigg\rvert^2
    \end{equation*}
    
    Next, let $\mathcal{C'_G}$ be the circuit that computes the first $t$ layers of \scrg' exactly using $t\log(w) + \log(n) + 1$ qubits and $2t$ queries.
    Then, the final state of this circuit will be
    \begin{equation*}
        \ket{\psi_t^C} = \sum_{i\in [w]}\gamma_{i}\ket{\chi_i}\ket{i}\ket{0}\ket{0}
    \end{equation*}
    for some normalized states $\ket{\chi_i}$ of $(t-1)\log(w)$ qubits.
    Now, append a register of $\log(w)$ qubits initialized to $\ket{0}$ to the circuit $\mathcal{C'_G}$.
    For ease of computation, we append it in between the current second and the third register. So, the new state of the circuit will be
    \begin{equation*}
        \ket{\psi_t^C} = \sum_{j\in [w]}\gamma_{i}\ket{\chi_i}\ket{i}\ket{0}\ket{0}\ket{0} = R_1R_2R_3R_4R_5\text{~(say)}.
    \end{equation*}

    Now, we simulate the final step of \scrg as below:
    \begin{align*}
        \ket{\psi_t^C} &= \sum_{j\in [w]}\gamma_{i}\ket{\chi_i}\ket{i}\ket{0}\ket{0}\ket{0}\\
        &\xrightarrow{CX(R_2, R_3)} \sum_{i\in [w]} \gamma_i \ket{\chi_j}\ket{i}\ket{i}\ket{0}\ket{0}\\
        &\xrightarrow{\widehat{U}_L(R_2, R_4)} \sum_{i\in [w]} \gamma_i \ket{\chi_j}\ket{i}\ket{i}\ket{L(v_{0,i})}\ket{0}\\
        &\xrightarrow{O_X(R_4, R_5)} \sum_{i\in [w]} \gamma_i \ket{\chi_j}\ket{i}\ket{i}\ket{L(v_{0,j})}\ket{X_{L(v_{0,j})}}\\
        &\xrightarrow{C-U_{t}(R_2, R_3, R_5)} \sum_{i\in [w]} \gamma_i \ket{\chi_j}\ket{i}\Big(\sum_{j\in [w]} \beta_{i,j,X_i}^{(t)}\ket{j}\Big)\ket{L(v_{0,j})}\ket{X_{L(v_{0,j})}}\\
        &\xrightarrow{O_X(R_4, R_5)} \sum_{i\in [w]} \gamma_i \ket{\chi_j}\ket{i}\Big(\sum_{j\in [w]} \beta_{i,j,X_i}^{(t)}\ket{j}\Big)\ket{L(v_{0,j})}\ket{0}\\
        &\xrightarrow{\widehat{U}_L(R_2, R_4)} \sum_{i\in [w]} \gamma_i \ket{\chi_j}\ket{i}\Big(\sum_{j\in [w]} \beta_{i,j,X_i}^{(t)}\ket{j}\Big)\ket{0}\ket{0}\\
        &= \sum_{j\in [w]} \Big(\sum_{i\in [w]} \gamma_i \beta_{i,j,X_i}^{(t)}\ket{\chi_j}\ket{i}\Big)\ket{j}\ket{0}\ket{0} = \ket{\psi_f}\text{~(say)}\\
    \end{align*}

    Similar to the length $1$ case, $C-U_{t}$ is defined as
    \begin{equation*}
        C-U_{t} = \sum_{i\in [w]}\ket{i}\bra{i} \otimes \Big[\sum_{j\in [w]} \beta_{i,j,0}^{(t)}\ket{j}\bra{i} \otimes \ket{0}\bra{0} + \sum_{j\in [w]} \beta_{i,j,0}^{(t)}\ket{j}\bra{i} \otimes \ket{1}\bra{1}\Big].
    \end{equation*}
    Then finally on measuring the third register, we get that the probability of accepting an input $X$ is
    \begin{equation*}
        Pr_{\mathcal{C'_G}}[accept] = \sum_{v_j\in F'} \Bigg\lvert \sum_{i\in [w]} \gamma_i \beta_{i,j,X_i}^{(t)}\ket{i}\Bigg\rvert^2 = \sum_{v_j\in F'} \Bigg\lvert \sum_{i\in [w]} \gamma_i \beta_{i,j,X_i}^{(t)}\Bigg\rvert^2.
    \end{equation*}
    which equals the probability of accepting the input $X$ in \scrg'.
    So, we obtain a $((l+1)\log(w)+\log(n)+1)$-qubit $2l+2$-query circuit to simulate \scrg' exactly as required.    
\end{proof}



\end{document}